\documentclass[a4paper, letter, longauth]{aa}

\usepackage{natbib}
\usepackage{graphicx}
\usepackage{txfonts}
\usepackage[bookmarks=false, colorlinks=true, citecolor=blue, linkcolor=blue]{hyperref}
\usepackage{upgreek}
\usepackage{pdflscape}
\usepackage{mathrsfs}

\newcommand{\JWST}{{JWST}}
\def\micron{\hbox{\,$\upmu$m}}
\newcommand{\Lsun}{\hbox{$L_{\rm \odot}$}}
\newcommand{\Msun}{\hbox{$M_{\rm \odot}$}}

\newcommand\nodata{ ~$\cdots$~ }

\bibpunct{(}{)}{;}{a}{}{,}

\titlerunning{H$_3^+$ absorption and emission in local U\slash LIRGs with JWST\slash NIRSpec}
\authorrunning{Pereira-Santaella et al.}

\begin{document}

\title{H$_3^+$ absorption and emission in local U\slash LIRGs with JWST\slash NIRSpec: Evidence for high H$_2$ ionization rates}

\author{Miguel~Pereira-Santaella\inst{\ref{inst1}}
\and Eduardo~Gonz\'alez-Alfonso\inst{\ref{inst2}}
\and Ismael~Garc\'ia-Bernete\inst{\ref{inst3}, \ref{inst4}}
\and Fergus~R.~Donnan\inst{\ref{inst4}}
\and Miriam~G.~Santa-Maria\inst{\ref{inst5}}
\and Javier~R.~Goicoechea\inst{\ref{inst1}}
\and Isabella~Lamperti\inst{\ref{inst6},\ref{inst7}}
\and Michele~Perna\inst{\ref{inst8}}
\and Dimitra~Rigopoulou\inst{\ref{inst4},\ref{inst9}}
}

\institute{Instituto de F\'isica Fundamental, CSIC, Calle Serrano 123, 28006 Madrid, Spain \\
\email{miguel.pereira@iff.csic.es}\label{inst1}
\and
Universidad de Alcal\'a, Departamento de F\'isica y Matem\'aticas, Campus Universitario, 28871 Alcal\'a de Henares, Madrid, Spain\label{inst2}
\and
Centro de Astrobiolog\'ia (CAB), CSIC-INTA, Camino Bajo del Castillo s/n, E-28692 Villanueva de la Ca\~nada, Madrid, Spain\label{inst3}
\and
Department of Physics, University of Oxford, Keble Road, Oxford OX1 3RH, UK\label{inst4}
\and
Department of Astronomy, University of Florida, P.O. Box 112055, Gainesville, FL 32611, US \label{inst5}
\and
Dipartimento di Fisica e Astronomia, Università di Firenze, Via G. Sansone 1, 50019, Sesto F.no (Firenze), Italy\label{inst6}
\and
INAF - Osservatorio Astrofisco di Arcetri, largo E. Fermi 5, 50127 Firenze, Italy\label{inst7}
\and
Centro de Astrobiolog\'ia (CAB), CSIC-INTA, Ctra de Torrej\'on a Ajalvir, km 4, 28850, Torrej\'on de Ardoz, Madrid, Spain\label{inst8}
\and
School of Sciences, European University Cyprus, Diogenes street, Engomi, 1516 Nicosia, Cyprus\label{inst9}
}

\abstract{
We study the $3.4-4.4$\micron\ fundamental rovibrational band of H$_3^+$, a key tracer of the ionization of the molecular interstellar medium (ISM), in a sample of 12 local ($d$$<$400\,Mpc) ultra\slash luminous infrared galaxies (U\slash LIRGs) observed with JWST\slash NIRSpec.
The P-, Q-, and R-branches of the band are detected in 13 out of 20 analyzed regions within these U\slash LIRGs, which increases the number of extragalactic H$_3^+$ detections by a factor of 6.
For the first time in the ISM, the H$_3^+$ band is observed in emission in 3 of these regions. In the remaining 10 regions, the band is seen in absorption.
The absorptions are produced toward the $3.4-4.4$\micron\ hot dust continuum rather than toward the stellar continuum, indicating that they likely originate in clouds associated with the dust continuum source.
The H$_3^+$ band is undetected in Seyfert-like \hbox{U\slash LIRGs} where the mildly obscured X-ray radiation from the AGN might limit the abundance of this molecule. For the detections, the H$_3^+$ abundances, $N($H$_3^+$)\slash $N_{\rm H}$=\hbox{(0.5--5.5)$\times$10$^{-7}$}, imply relatively high ionization rates, $\zeta_{\rm H_2}$, between 3$\times$10$^{-16}$ and $>$4$\times$10$^{-15}$\,s$^{-1}$, which are likely associated with high-energy cosmic rays.
In half of the targets the absorptions are blue-shifted by 50--180\,km\,s$^{-1}$, which are lower than the molecular outflow velocities measured using other tracers such as OH\,119\micron\ or rotational CO lines. This suggests that H$_3^+$ traces gas close to the outflow launching sites before it has been fully accelerated.
We used nonlocal thermodynamic equilibrium models to investigate the physical conditions of these clouds. In 7 out of 10 objects, the H$_3^+$ excitation is consistent with inelastic collisions with H$_2$ in warm translucent molecular clouds ($T_{\rm kin}$$\sim$250--500\,K and $n({\rm H_2})$$\sim$10$^{2-3}$\,cm$^{-3}$).
In three objects, dominant infrared pumping excitation is required to explain the absorptions from the (3,0) and (2,1) levels of H$_3^+$ detected for the first time in the ISM.
}


\maketitle

\section{Introduction}\label{sec:intro}

Galaxies with high infrared (IR) luminosities ($L_{\rm IR}>$10$^{11.5}$\Lsun), known as luminous and ultra-luminous infrared IR galaxies (\hbox{U\slash LIRGs}), are mostly gas-rich major mergers at different evolutionary stages (e.g., \citealt{Hung2014}). They represent a key phase in the evolution of galaxies both locally and at high-$z$ (e.g., \citealt{RodriguezGomez2015}).
U\slash LIRGs host the strongest starbursts in the local Universe, with star-formation rates \hbox{$>$30--70}\,\Msun\,yr$^{-1}$, and many of them contain bright active galactic nuclei (AGN) as well (e.g., \citealt{Veilleux2009, Nardini2010}).
U\slash LIRGs also host massive molecular outflows with mass outflow rates up to 300\,\Msun\,yr$^{-1}$ \citep{GonzalezAlfonso2017, Lutz2020, Lamperti2022}, which are expected to significantly influence their evolution by depleting the gas available for star formation and for fueling the central black hole. Most of the activity of U\slash LIRGs takes place in compact ($d$$<$200\,pc; \citealt{BarcosMunoz2017, Pereira2021}) deeply dust-embedded cores ($N_{\rm H}$$>$10$^{24}$\,cm$^{-2}$; \citealt{GonzalezAlfonso2015, Falstad2021, GarciaBernete2022_CON, Donnan2023}). 
These environments show a rich molecular gas chemistry, especially in the most obscured cases (e.g., \citealt{GonzalezAlfonso2012, Costagliola2015, Gorski2023}).
Cosmic rays have been proposed as the primary driver of the ionization and chemistry in these objects, as UV photons are shielded at high column densities. Additionally, cosmic rays can influence the conditions for star formation by heating the cores of dense molecular clouds \citep{Papadopoulos2010, Padovani2020}.

In cosmic-ray dominated regions (CRDR), H$_3^+$ is a key molecule which initiates the interstellar chemistry by donating a proton to other atoms and molecules (e.g., \citealt{Oka2013}). The H$_3^+$ formation is closely linked to the H$_2$ ionization rate, $\zeta_{\rm H_2}$: after the ionization of an H$_2$ molecule, H$_2^+$
readily reacts with another H$_2$ molecule to form H$_3^+$. Its formation\slash destruction balance is relatively simple, particularly in diffuse clouds ($n({\rm H_2})$$\sim$10$^{2}$\,cm$^{-3}$; e.g., \citealt{Dalgarno2006}), so H$_3^+$ has been used to measure $\zeta_{\rm H_2}$ in Galactic regions. H$_3^+$ is also abundant in X-ray dominated regions (XDR) when the X-ray radiation field has been sufficiently attenuated \citep{Maloney1996}. Thus, for objects with an AGN, the H$_3^+$ abundance can be affected by the X-ray radiation too.

Galactic H$_3^+$ absorptions have been detected toward the Galactic Center (GC), dense molecular clouds, and diffuse clouds using ground based observations (e.g., \citealt{Geballe1996, McCall2002, Goto2008, Gibb2010}). In these environments, $\zeta_{\rm H_2}$ varies between $\sim$10$^{-17}$\,s$^{-1}$ in dense clouds \citep{McCall1999}, $\sim$10$^{-16}$\,s$^{-1}$ in diffuse clouds \citep{Indriolo2012}, and $\sim$10$^{-14}$\,s$^{-1}$ in the GC \citep{LePetit2016, Oka2019}.
Two extragalactic H$_3^+$ detections have been reported so far: in the ULIRG IRAS\,08572$+$3915\,NW (\citealt{Geballe2006}; which is also part of the sample studied here), and a much fainter detection in the Type 2 AGN NGC\,1068 \citep{Geballe2015}.

In this Letter, we analyze James Webb Space Telescope (JWST) Near IR Spectrograph (NIRSpec; \citealt{Jakobsen2022}) observations that cover the 3.4--4.4\micron\ spectral range where the H$_3^+$ fundamental rovibrational $\nu_2$ band lies.
We study the kinematics of the absorptions and measure the H$_3^+$ column densities.
We also estimate $\zeta_{\rm H_2}$ from the H$_3^+$ abundance and investigate the physical conditions of the clouds producing these absorptions using radiative transfer models. We used the spectroscopic parameters of H$_3^+$ from \citet{Mizus2017}.

\begin{figure}
\centering
\includegraphics[width=0.48\textwidth]{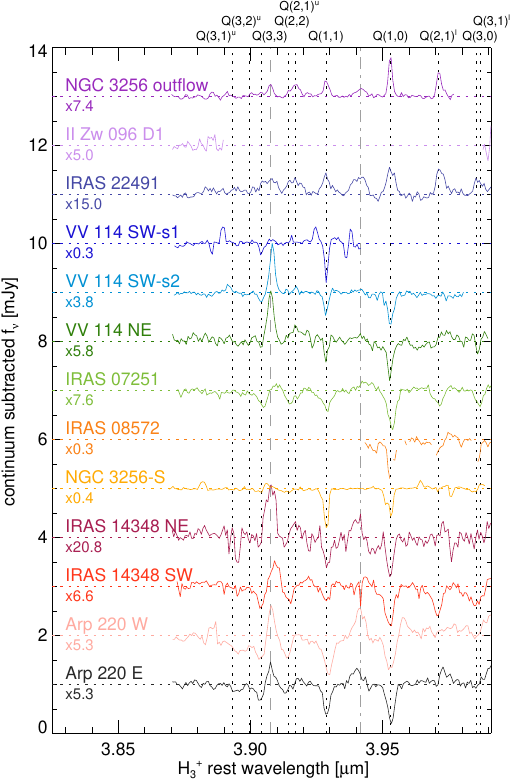}
\caption{Q-branch of the fundamental rovibrational $\nu_2$ band of H$_3^+$ (P- and R-branches are shown in Fig.~\ref{fig_h3p_rp_band}). The \JWST\slash NIRSpec spectra are continuum subtracted (Sect.~\ref{ss:columns}) and scaled. The H$_3^+$ transitions are labeled at the top of the panel and indicated by the dotted black vertical lines. Dashed and dot-dashed gray vertical lines indicate transitions of H$_2$ and \ion{H}{I}, respectively. The number below the region name is the scaling factor applied. In this figure, the rest frame is defined by the velocity of the H$_3^+$ features.
We note that the \JWST\slash NIRSpec spectra have a $\sim$0.1\micron\ gap centered around 4.05--4.15\micron\ that partially affects the Q-branch of some of these regions.
For IRAS\,08572, we masked spectral channels with highly uncertain flux values.
\label{fig_h3p_q_band}}
\end{figure}

\section{Analysis and results}\label{s:results}

We extracted the high resolution ($R$$\sim$1900--3600) JWST\slash NIRSpec spectra of 20 regions (nuclei and bright near-IR clumps) in 12 local (39--400\,Mpc) U\slash LIRGs with $L_{\rm IR}$=10$^{11.6-12.5}$\Lsun. All these U\slash LIRGs are interacting systems at different merger stages. The spectra were extracted using 0\farcs 40 diameter apertures (80--800\,pc depending on the distance). More details on the sample and the data reduction are presented in Appendix~\ref{apx:data} and Table~\ref{tbl_sample}.

Figure~\ref{fig_h3p_q_band} shows the Q-branch of the fundamental $\nu_2$ band of H$_3^+$ in these objects (the P- and R- branches are shown in Fig.~\ref{fig_h3p_rp_band}). The H$_3^+$ band is detected in 13 out of the 20 sources analyzed (the non detections are shown in Fig.~\ref{fig_h3p_band_nodet}). It is seen in absorption in ten targets and, for the first time, this band is detected in emission from gas in the interstellar medium (ISM) in the nuclei of two U/LIRGs and the molecular outflow of another target. Previous detections of H$_3^+$ emission were limited to the giant gas and ice planets of the Solar system (e.g., \citealt{Drossart1989, Trafton1993}). In this Letter, we focus on the targets where the band is detected in absorption. The analysis of the emission bands will be presented in a future work (Pereira-Santaella et al. in prep.).

\subsection{H$_3^+$ spectroscopy and Galactic observations}

\begin{figure}
\centering
\includegraphics[width=0.33\textwidth]{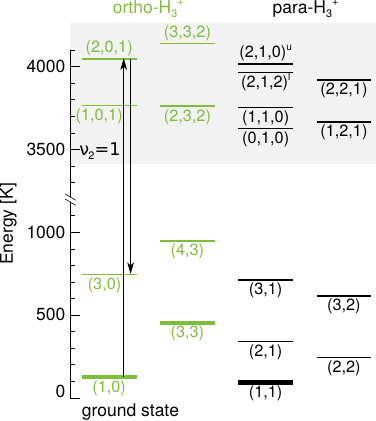}
\caption{Energy level diagram of H$_3^+$. The gray shaded area marks the $v_2=1$ levels.
Ortho-H$_3^+$ levels are in green and para-H$_3^+$ in black. The metastable and ground rotational levels are indicated by thicker lines. For the ground state, the quantum numbers of each level are ($J$,$K$) and for the $v_2=1$ levels, ($J$,$G$,$K$). The detected transitions connect the ground and the $v_2=1$ states. The arrows show how the (3,0) level can be populated by IR pumping through the R(1,0) and P(3,0) transitions.
\label{fig_levels}}
\end{figure}

The energy level diagram of H$_3^+$ is shown in Fig.~\ref{fig_levels}.
There are two H$_3^+$ spin isomers depending on the total nuclear spin $I$: ortho-H$_3^+$ with $I$=3\slash 2 and quantum number $K$=$3n$; and para-H$_3^+$ with $I$=1\slash 2 and $K$=$3n\pm1$ (green and black levels in Fig.~\ref{fig_levels}, respectively). Due to the selection rules, the (3,3) level of the ground state is metastable and cannot decay radiatively to the lowest ortho-H$_3^+$ level (1,0) (see e.g., \citealt{Oka2013, Miller2020} for a detailed description of the H$_3^+$ spectroscopy). Therefore, H$_3^+$ molecules tend to accumulate in the (1,1), (1,0), and (3,3) levels in the conditions of the ISM.
Actually, Galactic observations of H$_3^+$, in both diffuse and dense ($n({\rm H_2})$$\sim$10$^{4-5}$\,cm$^{-3}$) clouds, detect only the (1,1) and (1,0) absorptions \citep{McCall1999, Indriolo2012}, whereas the higher excitation (3,3) absorptions are observed only toward the GC. A (2,2) absorption has been uniquely detected along a line of sight (GC~IRS\,3) toward the circumnuclear disk (CND) of the GC \citep{Goto2008, Goto2014, Oka2019}.

\subsection{H$_3^+$ kinematics: outflows and inflows}

\begin{figure}[t]
\centering
\includegraphics[width=0.36\textwidth]{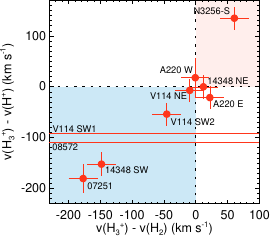}
\caption{Comparison between the velocity measured for the H$_3^+$ absorptions and that of H$_2$ ($x$ axis) and H recombination lines ($y$ axis). For two objects, no H$_2$ lines are available, so the horizontal line marks the   H$_3^+$ velocity relative to that of the H recombination lines. The shaded blue and red areas indicate blue and red velocity shifts, respectively.
\label{fig_kin}}
\end{figure}

The absorptions were fitted using Gaussian profiles. As in \citet{Pereira2024CO}, we tied the line of sight velocity of all the H$_3^+$ transitions to a common value. We also tied the velocity dispersion of the transitions of the same branch, but allowed it to vary between the three branches to account for the variation of the resolving power of NIRSpec (see Appendix~\ref{apx:data}).

In 50\% of the sample (5 objects), the H$_3^+$ absorptions are blue-shifted 50--180\,km\,s$^{-1}$ relative to the molecular and ionized gas traced by the high-$J$ pure rotational H$_2$ and H recombination lines observed by NIRSpec. This indicates that the clouds with H$_3^+$ are outflowing (Fig.~\ref{fig_kin}). These H$_3^+$ outflow velocities are lower than those measured using the OH\,119\micron\ and the CO(2--1)\,230.5\,GHz lines for these targets ($\sim$300--500\,km\,s$^{-1}$; \citealt{Veilleux2013, Lamperti2022}). Thus, it is possible that H$_3^+$ traces clouds close to the outflow launching sites before the gas has been fully accelerated (e.g., \citealt{GonzalezAlfonso2017, Pereira2020}).
Only in southern nucleus of NGC\,3256, the H$_3^+$ absorptions are red-shifted by $\sim$70--140\,km\,s$^{-1}$ suggesting an inflow.

\subsection{H$_3^+$ column densities and location of the clouds}\label{ss:columns}

We estimated the column density of each H$_3^+$ level using the standard relation for optically thin lines:
\begin{equation}
N_l = \frac{8 \pi c }{A_{ul} \lambda^4} \frac{g_l}{g_u} W_\lambda,
\label{eq_nl}
\end{equation}
where, $W_\lambda$ is the equivalent width (EW) in wavelength units of the absorptions, $N_l$ the column density of the lower level, $\lambda$ the wavelength of the transition, $c$ the speed of light, $A_{\rm ul}$ the Einstein A-coefficient of the transition, and $g_l$ and $g_u$ the statistical weights of the lower and upper levels, respectively.

In order to measure $W_\lambda$, we used a spline-interpolated baseline to estimate the continuum level.
We note that the EW of the P(2,2) and P(2,1) transitions (affected by the 4.27\micron\ CO$_2$ ice and gas absorptions) and the R(3,3)$^l$ and R(3,3)$^u$ transitions (affected by the 3.4--3.6\micron\ stellar continuum features and PAH aliphatic bands) are relatively uncertain.

We find that, for the same level, the column densities derived from the shorter wavelength branches are lower than those derived from longer wavelength branches.
In particular, for the (1,1) and (3,3) levels, which show absorptions in two or three branches, the R-branch (3.4--3.7\micron) columns are, on average, $\sim$1.5 times lower than those derived from the Q-branch (3.9--4.0\micron), and the latter are $\sim$1.3 times lower than those derived from the P-branch (4.2--4.4\micron).

In these objects the stellar continuum dominates the near-IR continuum up to $\sim$3.5--3.9\micron, while the hot dust continuum dominates the spectra at longer wavelengths \citep{Donnan2024}.
Thus, the column density discrepancies can be explained if the H$_3^+$ absorptions are primarily produced toward the compact hot dust continuum ($r$$<$20\,pc in some of these objects; \citealt{Rich2023, GarciaBernete2024_SODA, Inami2022}), whereas the more spatially extended stellar continuum, not affected by the H$_3^+$ absorptions, reduces the observed EW at shorter wavelengths.
This also implies that the spatial extent of the H$_3^+$ clouds would be more similar to that of the compact dust continuum than to the extended stellar continuum.
The non detection of the R(3,3)$^u$ line at 3.43\micron\ (where the stellar continuum dominates) in 9 out of 10 regions further corroborates the proposed explanation. R(3,3)$^u$ should have an \hbox{EW$>$2--10} times higher than the derived upper limits based on the P(3,3) absorption at 4.35\micron.

\section{Discussion}

\begin{figure*}[ht]
\centering
\includegraphics[width=\textwidth]{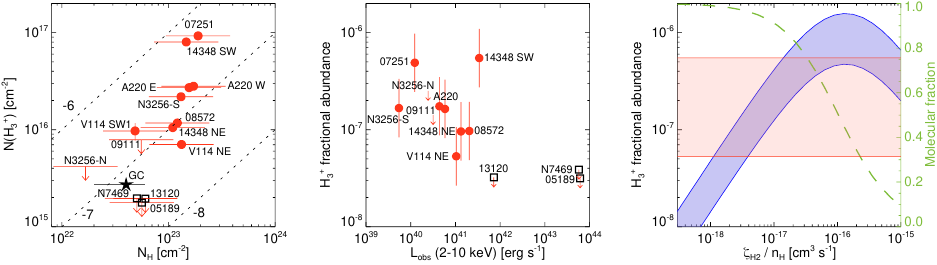}
\caption{Left panel: H$_3^+$ vs. H column densities. The H column density corresponds to the obscuring material of the $\sim$4\micron\ continuum derived from the continuum differential extinction model (Sect.~\ref{s:ion_rate}). The dashed lines indicate constant H$_3^+$ fractional abundances between 10$^{-6}$ and 10$^{-8}$ in 1\,dex steps. 
The red circles are the observed U\slash LIRGs. The black star corresponds to the average columns in the GC from  \citet{Goto2008} and \citet{Oka2019}. Upper limits for Seyfert type AGN are indicated by the empty black squares.
Middle panel: H$_3^+$ fractional abundance as function of the observed 2--10\,keV X-ray luminosity.
Right panel: The blue shaded area is the predicted H$_3^+$ fractional abundance (left $y$-axis) as function of $\zeta_{\rm H_2}$\slash $n_{\rm H}$ estimated using Eq.~2 from \citet{Neufeld2017} for $x_{\rm e}$ in the range (1.5--5)$\times$10$^{-4}$ (Sect.~\ref{s:ion_rate}). The horizontal red shaded area indicates the range of H$_3^+$ fractional abundances measured in these U\slash LIRGs. The dashed green line (right $y$-axis) is $f_{\rm H_2}$ as function of $\zeta_{\rm H_2}$\slash $n_{\rm H}$ using Eq.~2 from \citet{GonzalezAlfonso2013}.
\label{fig_h3p_abund}}
\end{figure*}

\subsection{H$_2$ Ionization rate: Cosmic rays and X-rays}\label{s:ion_rate}

For low $\zeta_{\rm H_2}$\slash $n_{\rm H}$ ($<$10$^{-17}$\,cm$^3$\,s$^{-1}$), where $n_{\rm H}$=$n$(H) + 2$n$(H$_2$), the H$_3^+$ abundance is proportional to $\zeta_{\rm H_2}$\slash $n_{\rm H}$. For higher $\zeta_{\rm H_2}$\slash $n_{\rm H}$, the molecular fraction, $f_{\rm H_2}$=2$n$(H$_2$)\slash $n_{\rm H}$, decreases, reducing the formation rate of H$_3^+$, and the increased abundance of free electrons, $x_{\rm e}$, enhances the dissociative recombinations of H$_3^+$. Hence, in this limit, the H$_3^+$ abundance decreases for increasing $\zeta_{\rm H_2}$\slash $n_{\rm H}$ in both XDR and CRDR (see \citealt{Maloney1996, LePetit2016, Neufeld2017}).

Thus, as a first step to estimate $\zeta_{\rm H_2}$, we calculated the H$_3^+$ fractional abundance, $N({\rm H_3^+})$\slash $N_{\rm H}$. We estimated $N_{\rm H}$ from the extinction affecting the 3.4--4.4\micron\ continuum where the H$_3^+$ absorptions are detected. This is justified since the H$_3^+$ absorptions are produced in clouds close to the continuum source (Sect.~\ref{ss:ratios}), which are likely the same clouds that extinguish this 3.4--4.4\micron\ continuum.
We used the method presented by \citet{Donnan2024} to model the differential extinction of the \hbox{3--28\micron} continuum observed with NIRSpec and MIRI\slash MRS for these sources.
We obtain an extinction at $\sim$4\micron\ equivalent to $N_{\rm H}$ \hbox{(2--19)}$\times$10$^{22}$\,cm$^{-2}$ (Table~\ref{tbl_H3p_cols}).
This is approximately equivalent to an optical depth $\tau_{\rm 4\mu m}$ about $\sim$$0.4-5$\footnote{For a screen extinction law \hbox{$N_{\rm H}$\slash $\tau_{4 \mu m}$}$\sim$3.8$\times$10$^{22}$\,cm$^{-2}$ \citep{Bohlin1978, Chiar2006}.}.
Higher $\tau_{\rm 4\mu m}$ would make this 3.4--4.4\micron\ dust continuum too weak to be detected.
However, we note that the $N_{\rm H}$ affecting the longer wavelength continuum in these U\slash LIRGs is considerably larger due to differential extinction effects.

Figure~\ref{fig_h3p_abund} shows that, for the detections, the average H$_3^+$ abundance is $\sim$2$\times$10$^{-7}$, which is slightly higher than the average abundance in the GC, $\sim$0.7$\times$10$^{-7}$, \citep{Oka2019}.
Interestingly, the $N({\rm H_3^+})$ upper limits for the three Sy AGN, i.e., the less obscured AGN in the sample where high-ionization lines are detected, imply low H$_3^+$ abundances $<$4$\times$10$^{-8}$. This result is also consistent with the low H$_3^+$ equivalent width measured in the Sy~2 AGN NGC\,1068 \citep{Geballe2015}.
This can be explained if the X-ray radiation from the AGN induces a $\zeta_{\rm H_2}$ high enough to suppress the formation of H$_3^+$ by both decreasing $f_{\rm H_2}$ and increasing $x_{\rm e}$.
We show in the middle panel of Fig.~\ref{fig_h3p_abund} the relation between the observed 2--10\,keV X-ray luminosity, $L_{\rm obs}$(2--10\,keV), and the H$_3^+$ abundance. The three Sy AGN have the highest $L_{\rm obs}$(2--10\,keV) and the stringent upper limits on the H$_3^+$ abundance. This suggests that X-rays might be limiting the formation of H$_3^+$.
We also estimated $\zeta_{\rm H_2}$ due to X-rays combining Eqs.~12 and 15 from \citet{Maloney1996}. For an $N_{\rm H}$ of 10$^{23}$\,cm$^{-2}$, \hbox{$\zeta_{\rm H_2}$\slash s$^{-1}$}=1.9$\times$10$^{-14}$$F_{\rm x}$\slash (erg\,s$^{-1}$\,cm$^{-2}$), where $F_{\rm x}$ is the incident X-ray flux. Therefore, an observed X-ray luminosity of 10$^{43}$\,erg\,s$^{-1}$, similar to that of the brightest Sy AGN in the sample, at 50\,pc implies $\zeta_{\rm H_2}$$\sim$6$\times$10$^{-13}$\,s$^{-1}$, which for the average $n({\rm H_2})$ of 10$^{2-3}$\,cm$^{-3}$ (Sect.~\ref{ss:ratios}), can place these objects in the regime where the H$_3^+$ abundance decreases with increasing $\zeta_{\rm H_2}$ (right side of the right panel of Fig.~\ref{fig_h3p_abund}).

In this section, we use the observed instead of the intrinsic X-ray luminosities because the H$_3^+$ absorptions originate at the outer obscuring layers ($N_{\rm H}$$\sim$10$^{23}$\,cm$^{-2}$) in objects with a total $N_{\rm H}$ up to 10$^{25}$\,cm$^{-2}$, thus a significant part of the intrinsic X-ray emission has been likely absorbed by more internal gas and dust layers.

The right panel of Fig.~\ref{fig_h3p_abund} shows the predicted H$_3^+$ abundance as function of $\zeta_{\rm H_2}$\slash $n_{\rm H}$ derived using Eq.~2 from \citet{Neufeld2017} at $T$=300\,K. Following \citet{GonzalezAlfonso2013}, we assumed $x_{\rm e}$ in the range (1.5--5)$\times$10$^{-4}$ and $f_{\rm H_2}$ given by their Eq.~2. This simplified approximation ignores most of the H$_3^+$ chemical reactions, but captures the two key elements, $x_{\rm e}$ and $f_{\rm H_2}$, that determine its abundance in diffuse and translucent clouds \citep{Dalgarno2006, Shaw2021}.
From the H$_3^+$ fractional abundance of the U\slash LIRGs, (0.5--5.5)$\times$10$^{-7}$, we estimate a \hbox{$\zeta_{\rm H_2}$\slash $n_{\rm H}$} between 10$^{-18}$ and $>$1.3$\times$10$^{-17}$\,cm$^3$\,s$^{-1}$. 
The average $n({\rm H_2})$ is $\sim$10$^{2.5}$\,cm$^{-3}$ (Sect.~\ref{ss:ratios}), so the resulting $\zeta_{\rm H_2}$ are between $\sim$3$\times$10$^{-16}$ and $>$4$\times$10$^{-15}$\,s$^{-1}$.

The origin, either X-rays or cosmic rays, of the H$_3^+$ detections is difficult to constrain. The effects of both X-rays and cosmic rays on the gas are similar (e.g., \citealt{Glassgold2012, Wolfire2022}).
Here, we utilize the fact that cosmic rays can penetrate column densities much larger than X-rays.
Therefore, if X-ray radiation from an AGN is present in these U\slash LIRGs, it would be absorbed by the internal gas layers, preventing it from reaching the more external gas clouds.
This hypothesis is supported by observations showing that U\slash LIRG are underluminous in hard X-rays \citep{Imanishi2004c, Teng2015, Ricci2021}.
Similarly, low energy proton cosmic-rays ($<$10--100\,MeV) have stopping ranges below 10$^{24}$\,cm$^{-2}$ \citep{Padovani2018}, so they are likely absorbed by the internal gas layers too.
However, more energetic proton cosmic-rays (100--180\,MeV), with stopping ranges of 10$^{24-25}$\,cm$^{-2}$, can deposit their energy over large columns comparable to those of U\slash LIRGs.
Therefore, we consider that ionization by $>$100\,MeV cosmic rays is more plausible than X-ray ionization in these objects.
These cosmic rays are more energetic than those in our Galaxy, which have typical energies of 2--10\,MeV \citep{Indriolo2009}. Thus, depending on both the cosmic-ray energy spectrum and the total $N_{\rm H}$ of U\slash LIRGs, the intrinsic cosmic-ray luminosity can be $>$20 times\footnote{This factor is the ratio between the effective $\zeta_{\rm H_2}$ for $N_{\rm H}$=10$^{22}$\,cm$^{-2}$ and 10$^{25}$\,cm$^{-2}$ given by Eq.~F.1 of \citet{Padovani2018} for the cosmic-ray energy spectrum $\mathscr{H}$.
} higher than in the GC for the same $\zeta_{\rm H_2}$.

Other molecular ions like OH$^+$, H$_2$O$^+$, and H$_3$O$^+$, which can form via reactions with H$_3^+$, have been detected in U\slash LIRG hosting Seyfert AGN (including NGC\,7469 which is part of this sample; \citealt{vanderWerf2010, Aalto2011, Pereira2013, Pereira2014}). The low H$_3^+$ abundance in these AGN, supports the idea that these ions form through reaction chains involving H$^+$ and O$^+$, instead of H$_3^+$ (\citealt{Hollenbach2012, GonzalezAlfonso2013, GomezCarrasco2014}).

\subsection{Physical conditions of the H$_3^+$ clouds}\label{ss:ratios}

\begin{figure}[t]
\centering
\includegraphics[width=0.35\textwidth]{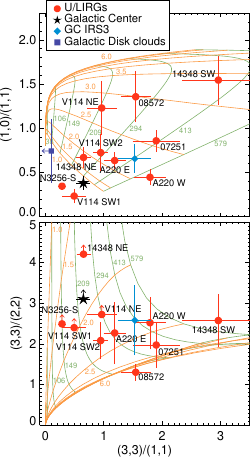}
\caption{Ratio between the  column densities of H$_3^+$ levels: (1,0)\slash (1,1) vs. (3,3)\slash (1,1) (top); and (3,3)\slash (2,2) vs. (3,3)\slash (1,1) (bottom). Symbols as in Fig.~\ref{fig_h3p_abund}.
The blue diamond corresponds to the excited GC~IRS\,3 line of sight toward the GC, and the dark blue square in the top panel to diffuse and dense molecular clouds in the Galactic disk. The background grids show the ratios predicted by NLTE models. The green (orange) lines mark the ratios for the temperature, in K, (density, in log cm$^{-3}$) indicated by the colored labels.
\label{fig_nlte_ratios}}
\end{figure}

Previous studies of the H$_3^+$ absorptions toward the GC indicate that H$_3^+$ traces a warm ($T$$\sim$200--500\,K) diffuse ($n_{\rm H}$$<$100\,cm$^{-3}$) gas phase with a high cosmic-ray ionization rate ($\zeta_{\rm cr}$$\sim$(1--10)$\times$10$^{-14}$\,s$^{-1}$; \citealt{Goto2008, LePetit2016, Oka2019}). In this section, we investigate if the conditions in these U\slash LIRGs are similar to those of the GC.

A major difference is that the observed H$_3^+$ column densities in these objects, $N({\rm H_3^+})$=(7--93)$\times$10$^{15}$\,cm$^{-2}$, are 2--30 times higher than in the GC.
The H$_3^+$ excitation is also different in some objects. Fig.~\ref{fig_nlte_ratios} shows column density ratios between various H$_3^+$ levels that can be used to characterize the physical conditions of these clouds (e.g., \citealt{Goto2008, LePetit2016}). As shown in the top panel of Fig.~\ref{fig_nlte_ratios}, contrary to Galactic diffuse and dense clouds, the (3,3) lines are detected in all these U\slash LIRGs. Thus, this ratio also suggests that the detections in U\slash LIRGs are not produced in extended diffuse gas halos (see also Sect.~\ref{ss:columns}).

In Fig.~\ref{fig_nlte_ratios}, we also plot a grid of nonlocal thermodynamic equilibrium (NLTE) models calculated using RADEX \citep{vanderTak2007} with the H$_3^+$-H$_2$ collisional coefficients from \citet{Oka2004}.
The RADEX grid covers H$_2$ densities, $\log n{(\rm H_2)}$\slash cm$^{-3}$, between 1.0 and 6.0, and kinetic temperatures, $T_{\rm kin}$, between 15 and 600\,K in logarithmic steps.

In three targets, VV\,114\,NE and SW-s1, and IRAS\,14348-NE, the (2,2) absorptions are not detected and the ratios are similar to those in the GC. Thus, for these nuclei, the gas temperature is 150--300\,K and $n({\rm H_2})$$<$100\,cm$^{-3}$, which are comparable to those of the GC (\citealt{Goto2008, LePetit2016, Oka2019}). We note that some of the observed (1,0)\slash (1,1) ratios are not well reproduced by these NLTE grids (top panel of Fig.~\ref{fig_nlte_ratios}). The formation of H$_3^+$ is highly exothermic (1.74\,eV$\equiv$20190\,K) and this can affect the observed populations of the (1,0) and (1,1) levels, which do not always reach the ortho-para thermal equilibrium  \citep{LeBourlot2024}.
Therefore, formation pumping might explain the low (1,0)\slash (1,1) ratio in some of these regions.
The remaining targets have clear (2,2) detections. For most objects, the \hbox{(3,3)\slash (2,2)} ratios (bottom panel of Fig.~\ref{fig_nlte_ratios}) are comparable, within the uncertainties, with that of the singular GC~IRS\,3 line of sight (\citealt{Goto2008, Goto2014}), and indicate warmer ($T_{\rm kin}$=250--500\,K) and denser ($\log n({\rm H_2})$$=$10$^{2-3}$\,cm$^{-3}$) clouds.

Some absorptions cannot be explained by these NLTE models. In particular, the absorptions from the (3,0) and (2,1) levels, detected in IRAS\,07251, IRAS\,08572, and IRAS\,14348-SW (Table~\ref{tbl_H3p_cols}), are 5--50 times stronger than those predicted by the best-fit NLTE models.
These absorptions, however, can be explained by IR pumping. When the IR pumping rate, which is proportional to the ambient IR radiation field $\phi_{\rm IR}$, is higher than the collisional rate, which is proportional to $n({\rm H_2})$, the (3,0) level, and similarly the (2,1) level, can be populated by the mechanism showed in Fig.~\ref{fig_levels} (see also Sect.~5.5 in \citealt{Goto2008}).
Therefore, it is likely that in these three objects, the $\phi_{\rm IR}$\slash $n({\rm H_2})$ ratio is high enough to make IR pumping the dominant excitation mechanism. We verified this scenario by creating a grid of full radiative transfer models, including the effects of IR pumping, which are able to reproduce these absorptions (Appendix~\ref{apx:models}).

\section{Summary and conclusions}\label{s:summary}

We analyzed the fundamental  \hbox{$3.4-4.4$\micron} rovibrational band of H$_3^+$ in the spectra of 20 regions (nuclei and bright clumps) selected in 12 local ($d$$\sim$39--400\,Mpc) U\slash LIRGs observed by JWST\slash NIRSpec. 
The H$_3^+$ band is detected in 13 out of 20 regions. It is seen in emission for the first time in the ISM in three of these regions, and in absorption in the remaining 10 regions. The main results from the analysis of the H$_3^+$ absorptions are the following:

\begin{enumerate}

\item \textit{H$_3^+$ clouds location.} The H$_3^+$ absorptions are primarily produced toward the hot dust continuum, which dominates the spectra at $>$3.5--3.9\micron, rather than toward the stellar continuum. 
Consequently, the clouds traced by these H$_3^+$  absorptions are likely associated with the compact ($r$$<$20\,pc in some cases) dust continuum source.

\item \textit{Non-detection of H$_3^+$ in Seyfert AGN.} The H$_3^+$ absorptions are undetected in the Seyfert-like U\slash LIRGs. These objects are among the less obscured ones in the sample. Thus they host mildly obscured AGN whose X-ray radiation might be limiting the abundance of H$_3^+$ by decreasing the molecular fraction $f_{\rm H_2}$ and increasing the free electron abundance $x_{\rm e}$.

\item \textit{Low velocity molecular outflows and inflows.} In five regions (50\% of the sample) the H$_3^+$ absorptions are blue-shifted by 50--180\,km\,s$^{-1}$ relative to the ionized and warm\slash hot molecular gas. These outflow velocities are lower than those measured using OH\,119\micron\ or rotational CO lines in these objects. This suggests that H$_3^+$ traces clouds closer to the outflow launching sites before it has been fully accelerated.
In one region the absorptions are red-shifted by 70--140\,km\,s$^{-1}$ suggesting inflowing gas.

\item \textit{H$_3^+$ column densities, abundances, and H$_2$ ionization rate.} The H$_3^+$ column densities are (0.7--9.3)$\times$10$^{16}$\,cm$^{-2}$ that correspond to fractional H$_3^+$ abundances of (0.5--5.5)$\times$10$^{-7}$.
Using a simple model for H$_3^+$ abundance in translucent clouds, we estimate $\zeta_{\rm H_2}$\slash $n_{\rm H}$ between 10$^{-18}$ and $>$1.3$\times$10$^{-17}$\,cm$^3$\,s$^{-1}$. For $n({\rm H_2})$$\sim$10$^{2.5}$\,cm$^{-3}$ (see below), $\zeta_{\rm H_2}$ are between 3$\times$10$^{-16}$ and $>$4$\times$10$^{-15}$\,s$^{-1}$. Energetic cosmic-rays, with $E$$>$100--180\,MeV, have stopping ranges of $N_{\rm H}$$\sim$10$^{24-25}$\,cm$^{-2}$, comparable to the total $N_{\rm H}$ of U\slash LIRGs, so they are a plausible origin for these $\zeta_{\rm H_2}$.

\item \textit{NLTE models and excitation of H$_3^+$.} We find that in most regions, 7 out of 10, the H$_3^+$ excitation is consistent with warm translucent molecular clouds ($T_{\rm kin}$$\sim$250--500\,K and $n({\rm H_2})$$\sim$10$^{2-3}$\,cm$^{-3}$) excited by inelastic collisions with H$_2$. This conditions are similar to those observed in the extreme line of sight GC IRS 3 toward the CND of the GC.
In three objects, absorptions from the (3,0) and (2,1) levels of o-H$_3^+$ and p-H$_3^+$, respectively, are detected for the first time in the ISM. These objects require IR pumping excitation to explain these excited absorptions.

\end{enumerate}

These results show the potential of JWST to detect the key molecule H$_3^+$ in extragalactic objects.
This will allow to constrain the ionization and the chemistry of the molecular ISM (i.e., the initial conditions for star-formation) using H$_3^+$ for the first time in large samples of galaxies.

\begin{acknowledgements}
We are grateful to Octavio Roncero for insightful discussions about H$_3^+$ and reading the manuscript.
We thank the referee, David Neufeld, for a careful reading of the manuscript and a constructive report.
The authors acknowledge the GOALS and the NIRSpec GTO teams for developing their observing programs.
MPS acknowledges support under grants RYC2021-033094-I and CNS2023-145506 funded by MCIN/AEI/10.13039/501100011033 and the European Union NextGenerationEU/PRTR.
MPS, EGA, MGSM, and JRG acknowledge funding support under grant PID2023-146667NB-I00 funded by the Spanish MCIN/AEI/10.13039/501100011033.
EGA acknowledges grants PID2019-105552RB-C4 and PID2022-137779OB-C41 funded by the Spanish MCIN/AEI/10.13039/501100011033.
IGB is supported by the Programa Atracci\'on de Talento Investigador ``C\'esar Nombela'' via grant 2023-T1/TEC-29030 funded by the Community of Madrid.
FD acknowledges support from STFC through studentship ST/W507726/1.
MGSM acknowledges support from the NSF under grant CAREER 2142300.
MP acknowledges grant PID2021-127718NB-I00 funded by the Spanish Ministry of Science and Innovation/State Agency of Research (MICIN/AEI/ 10.13039/501100011033).
DR acknowledges support from STFC through grants ST/S000488/1 and ST/W000903/1.

This work is based on observations made with the NASA/ESA/CSA James Webb Space Telescope. The data were obtained from the Mikulski Archive for Space Telescopes at the Space Telescope Science Institute, which is operated by the Association of Universities for Research in Astronomy, Inc., under NASA contract NAS 5-03127 for JWST; and from the European JWST archive (eJWST) operated by the ESAC Science Data Centre (ESDC) of the European Space Agency. These observations are associated with programs \#1267, \#1328 and \#3368.

\end{acknowledgements}

\bibliographystyle{aa}

\appendix

\onecolumn
\section{Sample and data reduction}\label{apx:data}

\begin{table}
\caption{Sample of local U\slash LIRGs}
\label{tbl_sample}
\centering
\begin{small}
\begin{tabular}{llcccccccc}
\hline \hline
\\
Object & Region & $D_{\rm L}$\,$^{(a)}$ & log\,$L_{\rm IR}\slash L_\odot$\,$^{(a)}$ & log\,$L^{\rm obs}_{\rm 2-10\,keV}$\,$^{(b)}$ & Ref.$^{(b)}$ & AGN\,$^{(c)}$ & Ref.$^{(c)}$ \\
& & (Mpc) & &  (erg\,s$^{-1}$) &  &  &  \\
\hline
VV\,114\,E & SW-s1 & 85.5 & 11.71  & \nodata & & No\slash Obs.? & D23, R23, B24 \\
         & SW-s2$^{(\dagger)}$  &  &  &  \nodata& \nodata & \nodata & B24, GA24 \\
         & NE &  &            & 41.02 & G06 & No\slash Obs.? & D23, R23, B24 \\ 

IRAS\,05189$-$2524 & nucleus & 188 & 12.10 & 43.79 &  I11 & Type 2 & P21 \\

IRAS\,07251$-$0248 & E nucleus & 400 & 12.45 & 40.09 &  I11, R21 & No\slash Obs. & P21, PS21 \\

IRAS\,08572$+$3915 & NW nucleus & 261 & 12.15 & 41.31 &  I11, R21 & LINER & V13 \\

IRAS\,09111$-$1007 & W nucleus & 242 & 12.05 & 40.50 &  I11 & LINER? & D97 \\

NGC\,3256 & N nucleus& 38.9 & 11.64  & 40.39 &  L02 & No & L00 \\ %
         & S nucleus &  &            & 39.73&  L02 &  No\slash Obs. & O15 \\
         & outflow$^{(\ast)}$ &  \nodata&  \nodata &  \nodata& \nodata & \nodata & \nodata\\

IRAS\,13120$-$5453 & nucleus & 136 & 12.27 & 41.86 & R21 & Type 2 & P21 \\

IRAS\,14348$-$1447 & NE nucleus & 375 & 12.41 & 41.12 & R21 & LINER?\slash Obs.? & P21, PS21 \\
                  & SW nucleus &     &       & 41.53 & R21 & LINER?\slash Obs. & P21, PS21 \\

Arp\,220 & W nucleus & 78 & 12.19 &  40.77& C02 & LINER & P20 \\ %
        & E nucleus &  &    & 40.64&  C02 & LINER & P20 \\ %

II\,Zw\,096 & D1 & 161 & 11.94 & 41.32 & I11 & No\slash Obs.? & I22, GB24\\  %
        & C0$^{(\ddagger)}$ &  \nodata&   \nodata &  \nodata & \nodata & \nodata & \nodata\\ %
        & D0$^{(\ddagger)}$ &  \nodata&    \nodata&  \nodata &  \nodata& \nodata & \nodata\\ %

IRAS\,22491$-$1808 & E nucleus & 352 & 12.23 & 40.65 & I11 & No\slash Obs. & P21, PS21 \\

NGC\,7469 & nucleus & 70.8 & 11.65 & 43.73 & B03, PS11 & Type 1 & O93, U22, A23, B24 \\ %
\hline
\end{tabular}
\end{small}
\tablefoot{
$^{(a)}$ Luminosity distance and 8-1000\micron\ IR luminosity from \citet{Armus09} and \citet{Pereira2021}. 
$^{(b)}$ Observed 2--10\,keV X-ray luminosity and reference.
$^{(c)}$ AGN classification based on optical spectroscopy and references. 
Obs. indicates dust obscured AGN based on indirect evidences from the mid-IR and sub-mm. ``?'' indicates inconclusive classification.
$^{(\dagger)}$ SW-s2 is located $\sim$100\,pc away from the nucleus SW-s1. The highly excited CO and H$_2$O mid-IR absorptions of this region were analyzed by \citet{GonzalezAlfonso2024} and \citet{Buiten2024}.
$^{(\ast)}$ This region encompasses the collimated molecular outflow launched by the southern nucleus of NGC\,3256. It is located $\sim$110\,pc away from the nucleus and corresponds to region N1 of \citet{Pereira2024CO}.
$^{(\ddagger)}$ These two regions of II\,Zw\,096 are bright star-forming clumps located 0.5--1\,kpc from the nucleus D1 \citep{Inami2022}. 
}
\tablebib{
(A23) \citealt{Armus2023};
(B03) \citealt{Blustin2003};
(B24) \citealt{Bianchin2024};
(B24) \citealt{Buiten2024};
(C02) \citealt{Clements2002};
(D23) \citealt{Donnan2023};
(D97) \citealt{Duc1997};
(G06) \citealt{Grimes2006};
(GA24) \citealt{GonzalezAlfonso2024};
(GB24) \citealt{GarciaBernete2024_SODA};
(I11) \citealt{Iwasawa2011};
(I22) \citealt{Inami2022};
(L00) \citealt{Lipari2000};
(L02) \citealt{Lira2002};
(O93) \citealt{Osterbrock1993};
(O15) \citealt{Ohyama2015};
(P20) \citealt{Perna2020};
(P21) \citealt{Perna2021};
(PS11) \citealt{Pereira2011};
(PS21) \citealt{Pereira2021};
(R21) \citealt{Ricci2021};
(R23) \citealt{Rich2023};
(U22) \citealt{U2022};
(V13) \citealt{Veilleux2013}.
}

\end{table}

We selected all the U\slash LIRGs with publicly available high spectral resolution NIRSpec and Mid-Infrared Instrument (MIRI) spectroscopic observations in the JWST archive by July 2024. We excluded few objects, mainly interacting systems whose nuclei are not clearly separated at the angular resolution of JWST\slash MIRI at $\sim$20\micron, which is needed for the continuum modeling of each nuclei.
The final sample consists of 12 systems (see Table~\ref{tbl_sample}) with data from the Director's Discretionary Early Release Science (DD-ERS) Program \#1328 (PI: L.~Armus and A.~Evans), the Large Program \#3368 (PI: L.~Armus and A.~Evans), and the Guaranteed Time Observations Program \#1267 (PI: D.~Dicken; Program lead: T.~B\"oker).

These objects were observed with the high spectral resolution NIRSpec grating G395H using the integral field unit (IFU) mode \citep{Boker2022}.
The G395H grating has a resolving power, $R$, between 1900 and 3600, and covers the spectral range between 2.87 and 5.27\micron\ (\JWST\ User Documentation).
Between $\sim$3.5\micron\ and $\sim$4.3\micron, the spectral range where the H$_3^+$ band is observed, $R$ varies between 2400 and 3000.
\JWST\slash MIRI observations of these objects were also available for all the bands of the Medium Resolution Spectrograph (MRS; \citealt{Wright2023, Argyriou2023}) between 4.9 and 28.1\micron\ with $R$=2300--3700 \citep{Labiano2021}.

For the data reduction, we used the \JWST\ calibration pipeline (version 1.12.4; \citealt{Bushouse_1_12_4}) and the context 1253. We followed the standard reduction recipe complemented by a number of custom steps to mitigate the effect of bad pixels and cosmic rays on the extracted spectra. Further details on the data reduction of the NIRSpec and MIRI\slash MRS data observations can be found in \citet{Pereira2022, Pereira2024CO} and \citet{GarciaBernete2022_PAH_MRS, GarciaBernete2024_ice}.

We extracted the spectra of 20 regions within these systems. Sixteen regions correspond to their nuclei, three regions are bright near-IR clumps with strong bursts of star-formation \citep{Inami2022, GarciaBernete2024_SODA, Buiten2024}, and one region samples the spatially resolved molecular outflow of NGC\,3256-S (\citealt{Pereira2024CO}; Table~\ref{tbl_sample}). We applied an aperture correction to the spectra similar to \citet{GarciaBernete2022}.

\section{Properties of the H$_3^+$ absorptions}

The EW of the detected H$_3^+$ absorptions and upper limits are listed in Tables~\ref{tbl_ew_r}, \ref{tbl_ew_q}, and \ref{tbl_ew_p}, for the R-, Q-, and P-branches, respectively.
Table~\ref{tbl_H3p_cols} shows the total H$_3^+$ column densities.

\begin{table}
\caption{EW of the H$_3^+$ R-branch transitions and derived column densities}
\label{tbl_ew_r}
\centering
\begin{small}
\begin{tabular}{lcccccccccccccccc}
\hline \hline
\\
Target  & R(3,3)$^u$ & $N$(3,3) & R(3,3)$^l$ & $N$(3,3) & R(1,0)\,$^{(a)}$ & $N$(1,0) & R(1,1)$^l$ & $N$(1,1) \\
  & 3.427\micron &  & 3.534\micron &  & 3.669\micron &  & 3.715\micron &  \\
  & (10$^{-5}$\,\micron) & (10$^{15}$\,cm$^{-2}$)  & (10$^{-5}$\,\micron) & (10$^{15}$\,cm$^{-2}$)  & (10$^{-5}$\,\micron) & (10$^{15}$\,cm$^{-2}$)  & (10$^{-5}$\,\micron) & (10$^{15}$\,cm$^{-2}$)  \\
\hline
VV\,114\,SW-s1 &  3.5 $\pm$ 0.8 & 3.5 $\pm$ 0.8 & 6.5 $\pm$ 0.7 & 2.3 $\pm$ 0.3 & 5.2 $\pm$ 0.6 & 1.3 $\pm$ 0.2 & 16.4 $\pm$ 0.6 & 7.6 $\pm$ 0.3 & \\
VV\,114\,SW-s2 &  $<$3.1 & $<$3.1 & 5.9 $\pm$ 1.4 & 2.1 $\pm$ 0.5 & 12.1 $\pm$ 1.1 & 3.1 $\pm$ 0.3 & 11.6 $\pm$ 1.0 & 5.4 $\pm$ 0.5 & \\
VV\,114\,NE &  $<$2.3 & $<$2.3 & $<$1.2 & $<$0.42 & 7.5 $\pm$ 1.1 & 1.9 $\pm$ 0.3 & 6.2 $\pm$ 1.1 & 2.9 $\pm$ 0.5 & \\
IRAS\,07251$^{(\ast)}$ &  $<$10.5 & $<$10.5 & $<$5.1 & $<$1.8 & 26.9 $\pm$ 3.8 & 6.8 $\pm$ 1.0 & \nodata & \nodata & \\
IRAS\,08572 &  $<$3.4 & $<$3.4 & \nodata & \nodata & 10.5 $\pm$ 0.6 & 2.7 $\pm$ 0.2 & 3.8 $\pm$ 0.6 & 1.8 $\pm$ 0.3 & \\
NGC\,3256-S &  $<$4.9 & $<$4.9 & 10.2 $\pm$ 1.2 & 3.6 $\pm$ 0.4 & 18.5 $\pm$ 0.6 & 4.7 $\pm$ 0.1 & 18.8 $\pm$ 0.5 & 8.7 $\pm$ 0.2 & \\
IRAS\,14348\,NE$^{(\ast)}$ &  $<$5.5 & $<$5.4 & $<$2.5 & $<$0.90 & 15.7 $\pm$ 2.1 & 4.0 $\pm$ 0.5 & \nodata & \nodata & \\
IRAS\,14348\,SW$^{(\ast)}$ &  $<$6.8 & $<$6.8 & $<$3.2 & $<$1.1 & 14.5 $\pm$ 2.6 & 3.7 $\pm$ 0.7 & $<$2.4 & $<$1.1 & \\
Arp\,220\,W &  $<$1.6 & $<$1.6 & 2.0 $\pm$ 0.7 & 0.70 $\pm$ 0.27 & 10.1 $\pm$ 0.7 & 2.6 $\pm$ 0.2 & 4.5 $\pm$ 0.7 & 2.1 $\pm$ 0.3 & \\
Arp\,220\,E &  $<$4.0 & $<$4.0 & 5.5 $\pm$ 1.6 & 2.0 $\pm$ 0.6 & 14.9 $\pm$ 1.1 & 3.8 $\pm$ 0.3 & 10.8 $\pm$ 1.0 & 5.0 $\pm$ 0.5 & \\
\hline\\
IRAS\,05189 &  $<$2.0 & $<$2.0 & $<$2.0 & $<$0.70 & $<$1.9 & $<$0.48 & $<$1.9 & $<$0.87 & \\
IRAS\,09111 &  $<$5.4 & $<$5.4 & $<$5.7 & $<$2.0 & $<$5.7 & $<$1.5 & $<$5.6 & $<$2.6 & \\
NGC\,3256-N &  $<$3.1 & $<$3.1 & $<$3.4 & $<$1.2 & $<$3.5 & $<$0.90 & $<$3.5 & $<$1.6 & \\
IRAS\,13120 &  $<$2.2 & $<$2.2 & $<$2.0 & $<$0.71 & $<$1.8 & $<$0.47 & $<$1.8 & $<$0.83 & \\
II\,Zw\,096\,C0 &  $<$3.6 & $<$3.6 & $<$3.8 & $<$1.3 & $<$3.7 & $<$0.95 & $<$3.8 & $<$1.7 & \\
II\,Zw\,096\,D0 &  $<$3.8 & $<$3.8 & $<$4.0 & $<$1.4 & $<$4.0 & $<$1.0 & $<$3.9 & $<$1.8 & \\
NGC\,7469 &  $<$2.0 & $<$2.0 & $<$2.0 & $<$0.70 & $<$1.9 & $<$0.49 & $<$1.9 & $<$0.88 & \\
\hline
\end{tabular}
\end{small}
\tablefoot{
$^{(a)}$ The R(1,0)\,3.669\micron\ and R(1,1)$^u$\,3.668\micron\ transitions are blended for these objects. To determine the $W_\lambda$ of R(1,0), the contribution of R(1,1)$^u$ was estimated from the $W_\lambda$ of R(1,1)$^l$.
$^{(\ast)}$ For these objects it was not possible to measure R(1,1)$^l$, thus the listed R(1,0) $W_\lambda$ value includes the contributions from both  R(1,0) and R(1,1)$^u$.
}
\end{table}

\begin{landscape}

\begin{table}
\caption{EW of the H$_3^+$ Q-branch transitions and derived column densities}
\label{tbl_ew_q}
\centering
\begin{small}
\begin{tabular}{lcccccccccccccccc}
\hline \hline
\\
Target  & Q(3,3) & $N$(3,3) & Q(2,2)\,$^{(a)}$ & $N$(2,2) & Q(1,1) & $N$(1,1) & Q(1,0) & $N$(1,0) & Q(2,1)$^l$ & $N$(2,1) & Q(3,0) & $N$(3,0) \\
  & 3.904\micron &  & 3.914\micron &  & 3.929\micron &  & 3.953\micron &  & 3.971\micron &  & 3.986\micron &  \\
  & (10$^{-5}$\,\micron) & (10$^{15}$\,cm$^{-2}$)  & (10$^{-5}$\,\micron) & (10$^{15}$\,cm$^{-2}$)  & (10$^{-5}$\,\micron) & (10$^{15}$\,cm$^{-2}$)  & (10$^{-5}$\,\micron) & (10$^{15}$\,cm$^{-2}$)  & (10$^{-5}$\,\micron) & (10$^{15}$\,cm$^{-2}$)  & (10$^{-5}$\,\micron) & (10$^{15}$\,cm$^{-2}$)  \\
\hline
VV\,114\,SW-s1 &  2.4 $\pm$ 0.5 & 2.3 $\pm$ 0.4 & $<$1.6 & $<$1.2 & 7.9 $\pm$ 0.5 & 3.8 $\pm$ 0.2 & \nodata & \nodata & \nodata & \nodata & \nodata & \nodata & \\
VV\,114\,SW-s2 &  5.1 $\pm$ 1.1 & 4.9 $\pm$ 1.0 & \nodata & \nodata & 10.9 $\pm$ 1.0 & 5.2 $\pm$ 0.5 & 18.8 $\pm$ 1.0 & 4.5 $\pm$ 0.2 & $<$3.4 & $<$1.2 & \nodata & \nodata & \\
VV\,114\,NE &  \nodata & \nodata & $<$1.1 & $<$0.78 & 4.6 $\pm$ 1.1 & 2.2 $\pm$ 0.5 & 11.2 $\pm$ 1.0 & 2.7 $\pm$ 0.3 & $<$1.0 & $<$0.38 & $<$3.6 & $<$0.89 & \\
IRAS\,07251 &  26.5 $\pm$ 4.3 & 25.4 $\pm$ 4.1 & 17.0 $\pm$ 4.2 & 12.2 $\pm$ 3.0 & 31.5 $\pm$ 4.1 & 15.2 $\pm$ 2.0 & 53.9 $\pm$ 3.8 & 13.0 $\pm$ 0.9 & 20.4 $\pm$ 3.7 & 7.4 $\pm$ 1.3 & 16.3 $\pm$ 3.6 & 4.0 $\pm$ 0.9 & \\
IRAS\,08572 &  \nodata & \nodata & \nodata & \nodata & \nodata & \nodata & 8.9 $\pm$ 1.3 & 2.1 $\pm$ 0.3 & 3.0 $\pm$ 1.2 & 1.1 $\pm$ 0.4 & \nodata & \nodata & \\
NGC\,3256-S &  \nodata & \nodata & $<$2.1 & $<$1.5 & 27.9 $\pm$ 0.7 & 13.4 $\pm$ 0.3 & 18.9 $\pm$ 0.7 & 4.6 $\pm$ 0.2 & $<$0.67 & $<$0.24 & \nodata & \nodata & \\
IRAS\,14348\,NE &  $<$1.3 & $<$1.3 & \nodata & \nodata & 9.4 $\pm$ 1.3 & 4.5 $\pm$ 0.6 & 12.5 $\pm$ 1.3 & 3.0 $\pm$ 0.3 & $<$1.3 & $<$0.46 & $<$1.2 & $<$0.30 & \\
IRAS\,14348\,SW &  12.0 $\pm$ 1.4 & 11.4 $\pm$ 1.4 & 6.2 $\pm$ 1.4 & 4.4 $\pm$ 1.0 & 8.0 $\pm$ 1.4 & 3.9 $\pm$ 0.7 & 24.7 $\pm$ 1.3 & 6.0 $\pm$ 0.3 & 17.6 $\pm$ 1.3 & 6.4 $\pm$ 0.5 & 12.5 $\pm$ 1.3 & 3.1 $\pm$ 0.3 & \\
Arp\,220\,W &  8.6 $\pm$ 2.1 & 8.3 $\pm$ 2.0 & 7.8 $\pm$ 2.1 & 5.6 $\pm$ 1.5 & 14.8 $\pm$ 2.0 & 7.1 $\pm$ 1.0 & 13.1 $\pm$ 2.0 & 3.2 $\pm$ 0.5 & $<$2.0 & $<$0.74 & $<$2.0 & $<$0.49 & \\
Arp\,220\,E &  8.4 $\pm$ 2.1 & 8.0 $\pm$ 2.0 & $<$2.1 & $<$1.5 & 16.9 $\pm$ 2.0 & 8.1 $\pm$ 1.0 & 21.3 $\pm$ 2.0 & 5.2 $\pm$ 0.5 & $<$2.0 & $<$0.72 & $<$1.9 & $<$0.47 & \\
\hline\\
IRAS\,05189 &  \nodata & \nodata & \nodata & \nodata & \nodata & \nodata & \nodata & \nodata & $<$1.6 & $<$0.58 & $<$1.6 & $<$0.40 & \\
IRAS\,09111 &  \nodata & \nodata & \nodata & \nodata & \nodata & \nodata & $<$7.4 & $<$1.8 & $<$7.1 & $<$2.6 & $<$7.2 & $<$1.8 & \\
NGC\,3256-N &  \nodata & \nodata & $<$3.2 & $<$2.3 & $<$3.2 & $<$1.5 & $<$3.2 & $<$0.76 & $<$3.2 & $<$1.2 & $<$3.2 & $<$0.78 & \\
IRAS\,13120 &  \nodata & \nodata & \nodata & \nodata & \nodata & \nodata & \nodata & \nodata & \nodata & \nodata & \nodata & \nodata & \\
II\,Zw\,096\,C0 &  \nodata & \nodata & \nodata & \nodata & \nodata & \nodata & \nodata & \nodata & \nodata & \nodata & \nodata & \nodata & \\
II\,Zw\,096\,D0 &  \nodata & \nodata & \nodata & \nodata & \nodata & \nodata & \nodata & \nodata & \nodata & \nodata & \nodata & \nodata & \\
NGC\,7469 &  $<$2.0 & $<$2.0 & \nodata & \nodata & \nodata & \nodata & $<$2.1 & $<$0.50 & $<$2.1 & $<$0.77 & $<$2.1 & $<$0.52 & \\
\hline
\end{tabular}
\end{small}
\tablefoot{
$^{(a)}$ The Q(2,2)\,3.914\micron\ and Q(2,1)$^u$\,3.916\micron\ transitions are blended for these objects. The Q(2,1)$^u$ absorption is $\sim$4 times fainter than the Q(2,1)$^l$, so only for IRAS\,07251 and IRAS\,14348\,SW, where the Q(2,1)$^l$ absorption is detected, we used its $W_\lambda$ to estimate the Q(2,1)$^u$ contribution.
}
\end{table}

\end{landscape}

\begin{table}
\caption{EW of the H$_3^+$ P-branch transitions and derived column densities}
\label{tbl_ew_p}
\centering
\begin{small}
\begin{tabular}{lcccccccccccccccc}
\hline \hline
\\
Target  & P(1,1)\,$^{(a)}$ & $N$(1,1) & P(2,2)\,$^{(b)}$ & $N$(2,2) & P(3,3) & $N$(3,3) & P(3,0) & $N$(3,0) \\
  & 4.070\micron &  & 4.204\micron &  & 4.350\micron &  & 4.355\micron &  \\
  & (10$^{-5}$\,\micron) & (10$^{15}$\,cm$^{-2}$)  & (10$^{-5}$\,\micron) & (10$^{15}$\,cm$^{-2}$)  & (10$^{-5}$\,\micron) & (10$^{15}$\,cm$^{-2}$)  & (10$^{-5}$\,\micron) & (10$^{15}$\,cm$^{-2}$)  \\
\hline
VV\,114\,SW-s1 &  7.9 $\pm$ 0.7 & 5.5 $\pm$ 0.5 & $<$0.67 & $<$0.25 & 8.3 $\pm$ 0.6 & 2.5 $\pm$ 0.2 & $<$0.65 & $<$0.48 & \\
VV\,114\,SW-s2 &  \nodata & \nodata & 6.4 $\pm$ 1.3 & 2.4 $\pm$ 0.5 & 17.0 $\pm$ 1.1 & 5.1 $\pm$ 0.3 & $<$3.8 & $<$2.8 & \\
VV\,114\,NE &  \nodata & \nodata & $<$1.0 & $<$0.37 & 7.1 $\pm$ 0.9 & 2.1 $\pm$ 0.3 & $<$0.89 & $<$0.66 & \\
IRAS\,07251 &  \nodata & \nodata & 45.7 $\pm$ 2.8 & 17.0 $\pm$ 1.0 & 107.5 $\pm$ 2.0 & 32.2 $\pm$ 0.6 & 18.3 $\pm$ 2.0 & 13.7 $\pm$ 1.5 & \\
IRAS\,08572 &  \nodata & \nodata & 5.7 $\pm$ 0.8 & 2.1 $\pm$ 0.3 & 9.2 $\pm$ 0.7 & 2.7 $\pm$ 0.2 & 2.1 $\pm$ 0.7 & 1.6 $\pm$ 0.5 & \\
NGC\,3256-S &  \nodata & \nodata & 6.5 $\pm$ 3.5 & 2.4 $\pm$ 1.3 & 12.7 $\pm$ 3.1 & 3.8 $\pm$ 0.9 & $<$3.1 & $<$2.3 & \\
IRAS\,14348\,NE &  \nodata & \nodata & $<$1.9 & $<$0.70 & 9.9 $\pm$ 1.2 & 3.0 $\pm$ 0.4 & $<$1.2 & $<$0.90 & \\
IRAS\,14348\,SW &  \nodata & \nodata & 24.8 $\pm$ 2.5 & 9.2 $\pm$ 0.9 & 70.3 $\pm$ 2.1 & 21.1 $\pm$ 0.6 & 12.0 $\pm$ 2.1 & 9.0 $\pm$ 1.6 & \\
Arp\,220\,W &  \nodata & \nodata & 13.6 $\pm$ 3.3 & 5.1 $\pm$ 1.2 & 42.5 $\pm$ 2.9 & 12.7 $\pm$ 0.9 & $<$2.9 & $<$2.2 & \\
Arp\,220\,E &  \nodata & \nodata & 11.4 $\pm$ 3.7 & 4.2 $\pm$ 1.4 & 32.1 $\pm$ 3.3 & 9.6 $\pm$ 1.0 & $<$3.4 & $<$2.5 & \\
\hline\\
IRAS\,05189 &  \nodata & \nodata & $<$1.6 & $<$0.59 & $<$1.5 & $<$0.46 & $<$1.5 & $<$1.1 & \\
IRAS\,09111 &  \nodata & \nodata & $<$12.2 & $<$4.5 & $<$12.8 & $<$3.8 & $<$12.9 & $<$9.6 & \\
NGC\,3256-N &  \nodata & \nodata & $<$3.7 & $<$1.4 & $<$3.8 & $<$1.1 & $<$3.8 & $<$2.8 & \\
IRAS\,13120 &  \nodata & \nodata & $<$2.1 & $<$0.77 & $<$2.0 & $<$0.61 & $<$2.0 & $<$1.5 & \\
II\,Zw\,096\,C0 &  \nodata & \nodata & $<$4.6 & $<$1.7 & $<$4.8 & $<$1.4 & $<$4.8 & $<$3.6 & \\
II\,Zw\,096\,D0 &  \nodata & \nodata & $<$3.3 & $<$1.2 & $<$3.2 & $<$0.94 & $<$3.1 & $<$2.3 & \\
NGC\,7469 &  \nodata & \nodata & $<$1.9 & $<$0.69 & $<$1.8 & $<$0.55 & $<$1.8 & $<$1.4 & \\

\hline
\end{tabular}
\end{small}
\tablefoot{
$^{(a)}$ The P(1,1) transition is blended with the H$_2$\,1-1\,S(13)\,4.068\micron. Thus, for most objects it is not possible to derive realiable measurements or upper limits.
$^{(b)}$ The P(2,2) transition is affected by the CO$_2$ ice and gas absorptions in some cases (e.g., NGC\,3256-S), so the derived EW and column densities for this transition are uncertain.
}
\end{table}

\begin{table}[h]
\caption{Column densities of H$_3^+$ and H}
\label{tbl_H3p_cols}
\centering
\begin{tiny}
\begin{tabular}{lcccccccccccccccc}
\hline \hline
\\
Target & $N$(H$_3^+$)$^{(a)}$ & $N$(1,1)$^{(b)}$ & $N$(1,0)$^{(b)}$ & $N$(2,2)$^{(b)}$ & $N$(2,1)$^{(b)}$ & $N$(3,3)$^{(b)}$ & $N$(3,0)$^{(b)}$ & $N_{\rm H}$$^{(c)}$ & $N$(H$_3^+$)\slash $N_{\rm H}$$^{(d)}$ \\
 & (10$^{15}$\,cm$^{-2}$)  & (10$^{15}$\,cm$^{-2}$)  & (10$^{15}$\,cm$^{-2}$)  & (10$^{15}$\,cm$^{-2}$)  & (10$^{15}$\,cm$^{-2}$)  & (10$^{15}$\,cm$^{-2}$)  & (10$^{15}$\,cm$^{-2}$)  & (10$^{22}$\,cm$^{-2}$) & (10$^{-7}$) \\
\hline
VV\,114\,SW-s1 &  9.7 $\pm$ 2.0 &  5.6 $\pm$ 1.9 &  1.3 $\pm$ 0.2 &  $<$1.2 & \nodata &  2.8 $\pm$ 0.7 & \nodata & 4.9 & 2.0 \\
VV\,114\,SW-s2 &  16.4 $\pm$ 1.3 &  5.2 $\pm$ 0.5 &  3.8 $\pm$ 1.0 &  2.4 $\pm$ 0.5 & \nodata &  5.0 $\pm$ 0.5 &  $<$2.8 & \nodata & \nodata \\
VV\,114\,NE &  7.0 $\pm$ 0.6 &  2.2 $\pm$ 0.5 &  2.7 $\pm$ 0.3 &  $<$0.78 &  $<$0.38 &  2.1 $\pm$ 0.3 &  $<$0.89 & 13.2 & 0.53 \\
IRAS\,07251 &  92.7 $\pm$ 6.6 &  15.2 $\pm$ 2.0 &  13.0 $\pm$ 0.9 &  14.6 $\pm$ 3.4 &  7.4 $\pm$ 1.3 &  28.8 $\pm$ 4.8 &  13.7 $\pm$ 1.5 & 18.9 & 4.9 \\
IRAS\,08572 &  11.7 $\pm$ 0.9 &  1.8 $\pm$ 0.3 &  2.4 $\pm$ 0.4 &  2.1 $\pm$ 0.3 &  1.1 $\pm$ 0.4 &  2.7 $\pm$ 0.2 &  1.6 $\pm$ 0.5 & 12.1 & 0.97 \\
NGC\,3256-S$^{(\ast)}$ &  21.8 $\pm$ 1.0 &  13.4 $\pm$ 0.3 &  4.6 $\pm$ 0.1 &  $<$1.5 &  $<$0.24 &  3.8 $\pm$ 0.9 &  $<$2.3 & 13.1 & 1.7 \\
IRAS\,14348\,NE &  10.5 $\pm$ 0.8 &  4.5 $\pm$ 0.6 &  3.0 $\pm$ 0.3 &  $<$0.70 &  $<$0.46 &  3.0 $\pm$ 0.4 &  $<$0.90 & 10.9 & 0.96 \\
IRAS\,14348\,SW$^{(\dagger)}$ &  80.2 $\pm$ 5.9 &  3.9 $\pm$ 0.7 &  6.0 $\pm$ 0.3 &  4.4 $\pm$ 1.0 &  6.4 $\pm$ 0.9 &  11.4 $\pm$ 1.4 &  3.1 $\pm$ 0.3 & 14.7 & 5.5 \\
Arp\,220\,W &  28.0 $\pm$ 1.9 &  7.1 $\pm$ 1.0 &  3.2 $\pm$ 0.5 &  5.1 $\pm$ 1.2 &  $<$0.74 &  12.7 $\pm$ 0.9 &  $<$0.49 & 17.2 & 1.6 \\
Arp\,220\,E &  27.1 $\pm$ 2.0 &  8.1 $\pm$ 1.0 &  5.2 $\pm$ 0.5 &  4.2 $\pm$ 1.4 &  $<$0.72 &  9.6 $\pm$ 1.0 &  $<$2.5 & 15.6 & 1.7 \\
\hline\\
IRAS\,05189 &  $<$1.8 &  $<$0.83 &  $<$0.48 &  $<$0.59 &  $<$0.58 &  $<$0.46 &  $<$0.40 & 5.6 &  $<$0.32 \\
IRAS\,09111 &  $<$7.9 &  $<$2.5 &  $<$1.6 &  $<$4.5 &  $<$2.6 &  $<$3.8 &  $<$5.7 & 5.6 &  $<$1.4 \\
NGC\,3256-N &  $<$4.2 &  $<$1.5 &  $<$0.83 &  $<$1.8 &  $<$1.2 &  $<$1.8 &  $<$1.8 & 1.7 &  $<$2.5 \\
IRAS\,13120 &  $<$2.0 &  $<$0.83 &  $<$0.47 &  $<$0.77 & \nodata &  $<$0.66 &  $<$1.5 & 6.1 &  $<$0.32 \\
II\,Zw\,096\,C0 &  $<$4.7 &  $<$1.7 &  $<$0.95 &  $<$1.7 & \nodata &  $<$2.1 &  $<$3.6 & \nodata & \nodata \\
II\,Zw\,096\,D0 &  $<$4.8 &  $<$1.7 &  $<$1.0 &  $<$1.2 & \nodata &  $<$2.1 &  $<$2.3 & \nodata & \nodata \\
NGC\,7469 &  $<$2.0 &  $<$0.84 &  $<$0.49 &  $<$0.69 &  $<$0.77 &  $<$0.62 &  $<$0.52 & 5.0 &  $<$0.39 \\

\hline
\end{tabular}
\end{tiny}
\tablefoot{
$^{(a)}$ Total column density of H$_3^+$ estimated by summing the population of the measured H$_3^+$ levels in this table.
We did not correct this total column density from the $\sim$10--50\%\ H$_3^+$ molecules in unobserved metastable levels, which depends on the physical conditions and excitation of the H$_3^+$ clouds (Appendix~\ref{apx:models} and \citealt{Oka2019}).
The upper limits are derived from the upper limits of the three lowest metastable levels of H$_3^+$: (1,1), (1,0), and (3,3).
$^{(b)}$ Adopted column density for the H$_3^+$ levels (1,1), (1,0), (2,2), (2,1), (3,3), and (3,0). We averaged the columns from the different branches, excluding, in most cases, those derived from the R-branch due to the dilution by the stellar continuum of the R-branch (Sect~\ref{ss:columns}). We also excluded some absorptions partially blended with emission lines in some of the objects (e.g., Q(3,3) in Arp\,220).
$^{(c)}$ H column density estimated from the extinction affecting the 3.4--4.4\micron\ continuum (see Sect.~\ref{s:ion_rate}; \citealt{Donnan2024}). The estimated uncertainty is $\sim$0.3\,dex.
$^{(d)}$ H$_3^+$ fractional abundance. The uncertainty is dominated by the 0.3\,dex uncertainty on $N_{\rm H}$.
$^{(\ast)}$ In NGC\,3256-S the P(2,2) transition is detected (Table~\ref{tbl_ew_p} and Fig.~\ref{fig_h3p_rp_band}), but its EW is affected by the broad CO$_2$ ice and gas absorptions. For this reason, we adopt the upper limit estimated from the Q(2,2) transition.
$^{(\dagger)}$ For IRAS\,14348\,SW, the P-branch column densities are 2.3$\pm$0.6 times larger than those from the Q-branch. Thus, we considered the Q-branch values for the column of the individual levels, which are used for population ratios, and obtained the total column density by multiplying their sum by 2.3.
}
\end{table}

\section{Spectra of the H$_3^+$ band}

\begin{figure}
\centering
\includegraphics[width=0.4\textwidth]{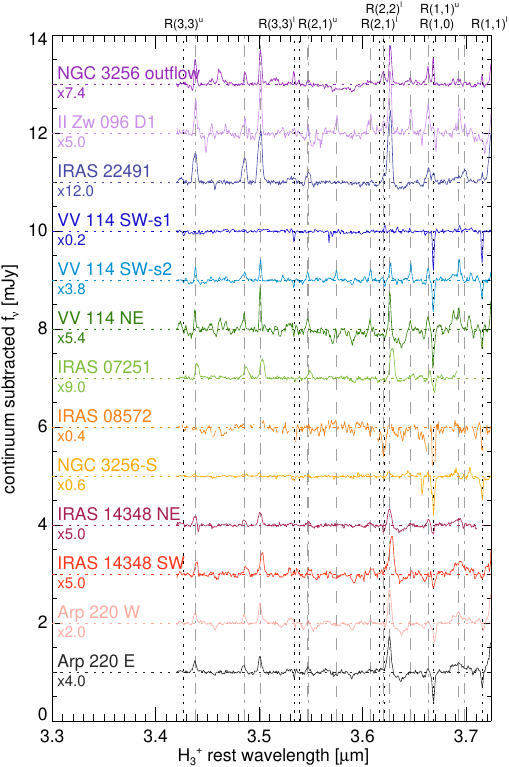}
\includegraphics[width=0.4\textwidth]{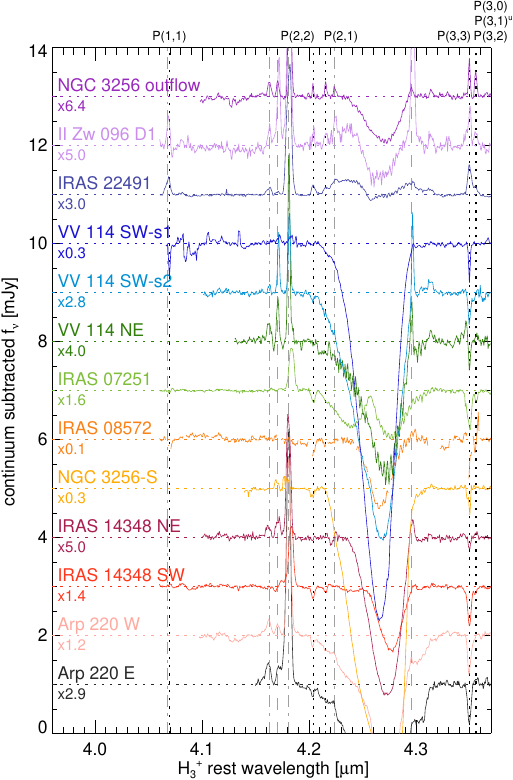}
\caption{Similar to Fig.~\ref{fig_h3p_q_band}, but for the R- and P-branches of H$_3^+$. In the P-branch panel, the \ion{He}{i} line at 4.296\micron\ is indicated by a dashed gray line.\label{fig_h3p_rp_band}}
\end{figure}

\begin{figure}
\centering
\includegraphics[width=0.4\textwidth]{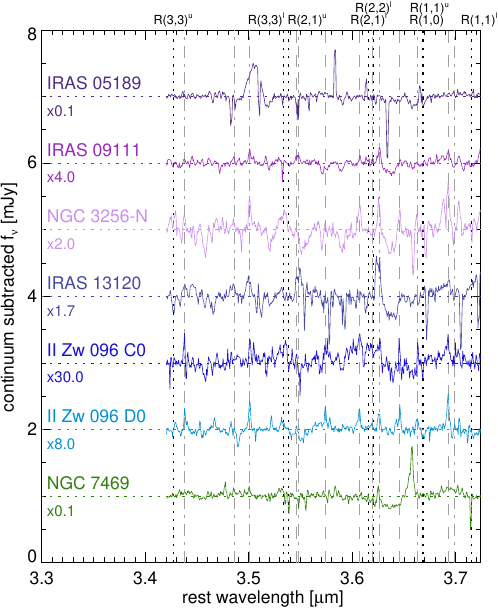}
\includegraphics[width=0.4\textwidth]{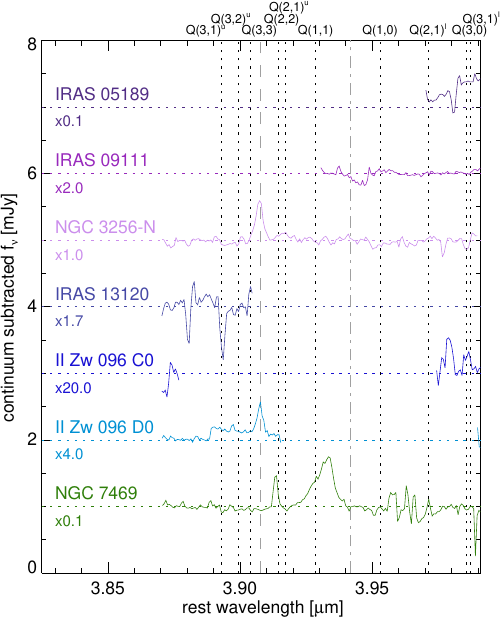}
\includegraphics[width=0.4\textwidth]{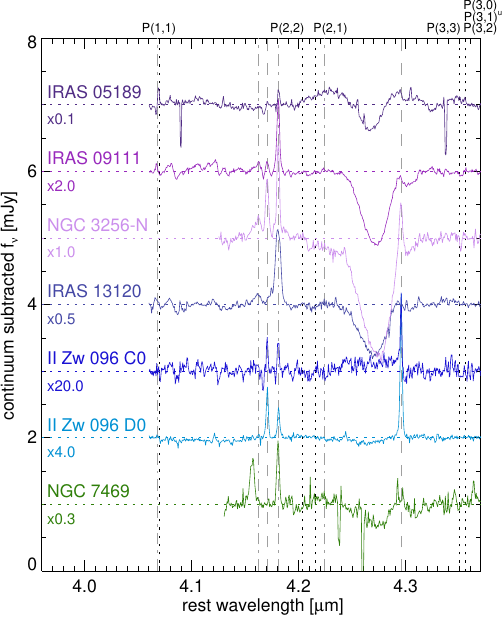}
\caption{Similar to Figs.~\ref{fig_h3p_q_band} and \ref{fig_h3p_rp_band}, but for the non-detections of the H$_3^+$ band. The rest frame is defined by the average velocity of the H$_2$ and H transitions.\label{fig_h3p_band_nodet}}
\end{figure}

Figure~\ref{fig_h3p_rp_band} presents the R- and P-branches of the detected H$_3^+$ band. The Q-branch is shown in Fig.~\ref{fig_h3p_q_band}.
Fig.~\ref{fig_h3p_band_nodet} shows the spectra of the regions where no H$_3^+$ is detected.

\clearpage
\section{Models of the H$_3^+$ absorptions including IR pumping}\label{apx:models}

\begin{figure}
\centering
\includegraphics[width=0.9\textwidth]{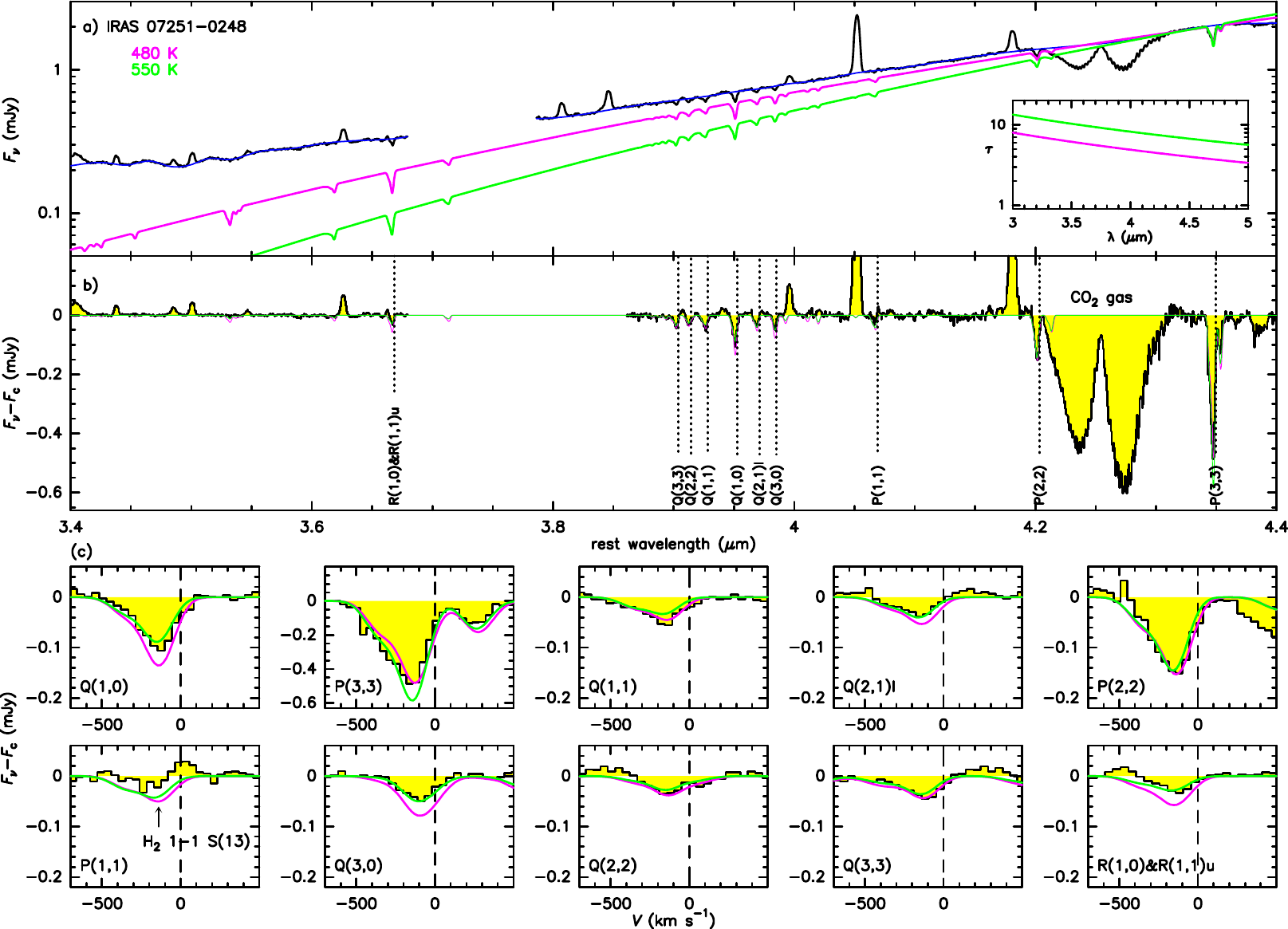}
\caption{
H$_3^+$ absorption in IRAS~07251$-$0248 and comparison with models. a) Observed $3.4-4.4$\,$\mu$m spectrum in IRAS~07251$-$0248 and adopted spline-interpolated baseline (in blue). The magenta and green lines show the modeled spectra, with the temperatures of the central source $T_c$ and the hot-dust optical depth ($\tau$ in the inset) also indicated with colors. b) The continuum-subtracted spectrum (yellow histogram) is compared with the predictions by the two models.
The spectra and models for individual lines are plotted in units of velocity in panel c. The vertical dashed line indicates the rest-frame zero velocity determined from the H$_2$ and \ion{H}{i} lines. The position of the H$_2$ $1-1$ S(13)\,4.068\micron\ line, which is blended with the P(1,1)\,4.070\micron\ line, is indicated by an arrow.
\label{fig_apx_mod_07251}}
\end{figure}

\begin{figure}
\centering
\includegraphics[width=0.9\textwidth]{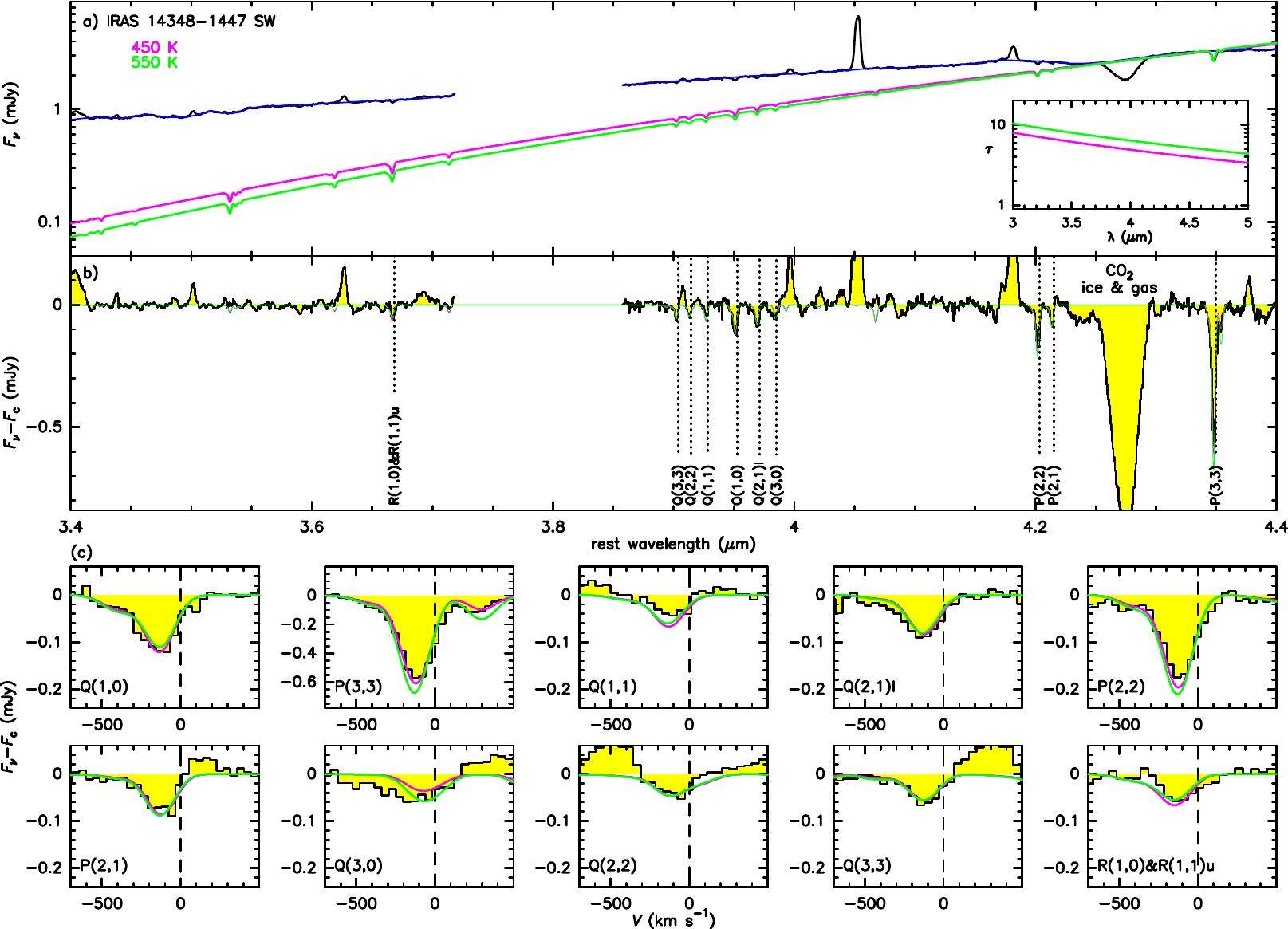}
\caption{Same as Fig.~\ref{fig_apx_mod_07251} but for IRAS~14348$-$1447~SW.\label{fig_apx_mod_14348}}
\end{figure}

We have modeled the H$_3^+$ absorption in IRAS\,07251$-$0248 and
IRAS\,14348$-$1447\,SW using the code described in \citet{GonzalezAlfonso1998}. The models
assume spherical symmetry with the statistical equilibrium of the
H$_3^+$ level populations calculated within a shell of gas
surrounding a mid-IR source of continuum radiation. The spectra
are computed towards the IR source excluding reemission from the
flanks. The effects of both radiative pumping by the IR source and
collisional excitation are included.
The varying parameters are the equivalent radius and temperature of the central
source ($R_c$ and $T_c$), the extinction by the dust mixed with the H$_3^+$
(assuming an extinction law with $\tau_{\lambda}\propto \lambda^{-1.7}$), the H$_3^+$ column density
of the shell ($N$(H$_3^+$)), the distance between the H$_3^+$ shell and
central source ($d$), the gas conditions ($n_{\mathrm{H_2}}$ and $T_{\rm kin}$),
and the gas velocity field. We assume that the dust mixed with the gas
is cold enough such that its reemission does not contribute to the
$<4.3\,\mu$m continuum.

Figures~\ref{fig_apx_mod_07251} and \ref{fig_apx_mod_14348} show the $3.4-4.4$\micron\ spectra of IRAS\,07251$-$0248 and
IRAS\,14348$-$1447\,SW, with the blue lines indicating the adopted spline-interpolated baselines.
As discussed in Sect.~\ref{ss:columns}, H$_3^+$ line absorption is progressively
diluted in the stellar continuum at shorter wavelengths, and hence we
attempt different spectral shapes for the hot-dust emission
characterized by $T_c$ and $\tau_{\mathrm{4\mu m}}$; $R_c$ is then determined by 
assuming negligible stellar continuum at 4.35\,$\mu$m (see \citealt{Donnan2024}).
We show in Figs.~\ref{fig_apx_mod_07251} and \ref{fig_apx_mod_14348} two models, A (in magenta) and B (in green), with
$T_c$, $\tau_{\mathrm{\lambda}}$ and the hot-dust emission indicated in panel a.
In these figures, the continuum-subtracted spectra in both sources is
compared with the predictions by the two models, and details of the observed
and modeled line shapes are shown in panels c. The physical parameters of the
models are listed in Table~\ref{apx:tabpar}.

These models show that, including IR pumping, it is possible to reproduce the absorptions from the (3,0) and (2,1) levels, which are difficult to explain just by collisional excitation with H$_2$ (Sect.~\ref{ss:ratios}).

\begin{table}
  \caption{Parameters for the models shown in Figs.~\ref{fig_apx_mod_07251} and \ref{fig_apx_mod_14348}}
  \label{apx:tabpar}
  \begin{center}
  \begin{tabular}{lcccccc}
    \hline
    \hline
    Source            &  Model  & $T_c$ & $R_c$ & $\tau_{\mathrm{4\mu m}}$ & $d$ & $N(\mathrm{H_3^+})$ \\
                      &         & (K)   &  (pc) &          & ($R_c$)       &  ($10^{17}$\,cm$^{-2}$) \\
                      
    \hline
    IRAS~07251$-$0248 & A (magenta) & 480 & 4.8 & 5.0 & 1.05 & 1.8 \\
                      & B (green)   & 550 & 10.9 &8.3 & 4.0 & 1.8 \\
    \hline
    IRAS~14348$-$1447~SW & A (magenta) & 450 & 5.9 &5.0 & 16 & 0.82 \\
                         & B (green)   & 550 & 5.7 &6.4 & 16 & 1.0 \\
    \hline
  \end{tabular}
  \end{center}
\end{table}

\end{document}